\documentclass[conference]{IEEEtran} \newcommand{\figsize}{0.9 \columnwidth}
\IEEEoverridecommandlockouts
\usepackage{graphicx,psfrag}
\usepackage[cmex10]{amsmath}
\usepackage{amsfonts,amssymb,latexsym,hyperref}
\usepackage[nospace]{cite}
\usepackage[usenames,dvipsnames]{color}
\usepackage{algorithm,algorithmic,setspace}
\usepackage{tikz}
\usetikzlibrary{shapes,arrows,patterns}

\makeatletter
\def\BState{\State\hskip-\ALG@thistlm}
\makeatother

 \newcommand{\putFrag}[4]{\begin{figure}[t]
                            \centering
                            #4
			    \includegraphics[width=#3]{figures/#1.eps}
            		    \caption{#2}
     			    \label{fig:#1}
                          \end{figure}
                          }

 \newcommand{\defn}{\triangleq}

 \newcommand{\tvec}[1]{\ensuremath{\Tilde{\boldsymbol{#1}}}}
 \newcommand{\ovec}[1]{\ensuremath{\Bar{\boldsymbol{#1}}}}
 \newcommand{\hvec}[1]{\ensuremath{\Hat{\boldsymbol{#1}}}}
    
 \renewcommand{\vec}[1]{\ensuremath{\boldsymbol{#1}}}
 \newcommand{\mat}[1]{\ensuremath{\begin{bmatrix}#1\end{bmatrix}}}
 \newcommand{\smallmat}[1]{\ensuremath{
        \left[\begin{smallmatrix}#1\end{smallmatrix}\right]}}

 \newcommand{\mc}[1]{\ensuremath{\mathcal{#1}}}

 \newcommand{\Real}{{\mathbb{R}}}
 \newcommand{\Complex}{{\mathbb{C}}}

 \newcommand{\tran}{^{\textsf{T}}}

 \newcommand*\dif{\mathop{}\!\mathrm{d}} 
 \newcommand{\bkt}[1]{{\langle #1 \rangle}}

 \DeclareMathOperator{\sgn}{sgn}
 
 \DeclareMathOperator{\E}{E}

 \DeclareMathOperator{\tr}{tr}

 \renewcommand{\eqref}[1]{(\ref{eq:#1})}
 
 \newcommand{\Figref}[1]{Figure~\ref{fig:#1}}
 
 \newcommand{\figref}[1]{Fig.~\ref{fig:#1}}

 \newcommand{\secref}[1]{Section~\ref{sec:#1}}

 \newcommand{\algref}[1]{Algorithm~\ref{alg:#1}}
 \newcommand{\lineref}[1]{line~\ref{line:#1}}

 \newfloat{myalgo}{tbhp}{mya}
  {\begin{myalgo}[#1]
        \centering
        \begin{minipage}{#2}
        \begin{algorithm}[H]}
  {\end{algorithm}
        \end{minipage}
        \end{myalgo}}

\newcommand{\pyzv}{p_{\vec{y}|\vec{z}}}
\newcommand{\pyz}{p_{y|z}}
\newcommand{\pxv}{p_{\vec{x}}}
\newcommand{\px}{p_{x}}
\newcommand{\pwv}{p_{\vec{w}}}

\newcommand{\map}{_{\text{\sf map}}}

\tikzstyle{block}=[rectangle,draw, fill=blue!20,
    minimum height=1cm, minimum width=1.5cm]
\tikzstyle{signal}=[coordinate,draw]

\begin{document}
\setlength{\arraycolsep}{0.4mm}
\title{Vector Approximate Message Passing for the Generalized Linear Model}
\author{
  \IEEEauthorblockN{
  Philip Schniter,\IEEEauthorrefmark{1}
         \thanks{\IEEEauthorrefmark{1}%
         Schniter acknowledges support from NSF grant 1527162; 
         Rangan from NSF grants 1302336, 1564142, and 1547332;
         and Fletcher from NSF grants 1254204 and 1564278 as well as
         ONR grant N00014-15-1-2677}%
  Sundeep Rangan,\IEEEauthorrefmark{2}
  and Alyson K.\ Fletcher\IEEEauthorrefmark{3}
  }
  \IEEEauthorblockA{
         \IEEEauthorrefmark{1}%
         Dept.\ of ECE, 
         The Ohio State University, Columbus, OH, 43210.  
         (Email: schniter.1@osu.edu)\\
         \IEEEauthorrefmark{2}%
         Dept.\ of Electrical and Computer Engineering,
         New York University, Brooklyn, NY, 11201.
         (Email: srangan@nyu.edu)\\
         \IEEEauthorrefmark{3}%
         Depts.\ of Statistics, Mathematics, and Electrical Engineering, 
         UCLA, Los Angeles, CA 90095. 
         (Email: akfletcher@ucla.edu)
  } 
}
\date{\today}
\maketitle

\begin{abstract}
The generalized linear model (GLM), where a random vector $\vec{x}$ is observed through a noisy, possibly nonlinear, function of a linear transform output $\vec{z}=\vec{Ax}$, arises in a range of applications such as 
robust regression, 
binary classification,
quantized compressed sensing, 
phase retrieval, 
photon-limited imaging,
and inference from neural spike trains.
When $\vec{A}$ is large and i.i.d.\ Gaussian,
the generalized approximate message passing (GAMP) algorithm is an efficient means of MAP or marginal inference,
and its performance can be rigorously characterized by a scalar state evolution.
For general $\vec{A}$, though, GAMP can misbehave. 
Damping and sequential-updating help to robustify GAMP, but their effects are limited.
Recently, a ``vector AMP'' (VAMP) algorithm was proposed for additive white Gaussian noise channels. 
VAMP extends AMP's guarantees from i.i.d.\ Gaussian $\vec{A}$ to the larger class of rotationally invariant $\vec{A}$.
In this paper, we show how VAMP can be extended to the GLM.
Numerical experiments show that the proposed GLM-VAMP is much more robust to ill-conditioning in $\vec{A}$ than damped GAMP.
\end{abstract}


\section{Introduction}              \label{sec:intro}

\begin{figure}
\centering
\begin{tikzpicture}[scale=1]
    \node (x) {$\vec{x} \sim \pxv$};
    \node [block,node distance=2.0cm]  (A)   [right of=x] {$\vec{A}$ };
    \node [block,node distance=2.25cm]  (pyz) [right of=A] {$\pyzv$}
        edge [<-] node[auto,swap] {$\vec{z}$} (A);
    \node [node distance=1.65cm] (y) [right of=pyz] {$\vec{y}$};

    \node [below of=x,font=\footnotesize] 
        {\parbox{1.5cm}{\centering Unknown input} };
    \node [below of=A,font=\footnotesize] 
        {\parbox{1.5cm}{\centering Linear transform} };
    \node [below of=pyz,text width=1.5cm,font=\footnotesize]
        {\parbox{1.5cm}{\centering Measurement channel} };
    \node [below of=y,text width=1.5cm,font=\footnotesize,xshift=0.0cm]
        {\parbox{2.0cm}{\centering Observed measurement} };
    \draw [->] (x) -- (A);
    \draw [->] (pyz) -- (y);
\end{tikzpicture}
\vspace{-2mm}
\caption{Generalized Linear Model (GLM): 
An unknown random vector $\vec{x}$
is observed through a linear transform $\vec{A}$ 
followed by a probabilistic measurement channel $\pyzv$,
yielding the measured vector $\vec{y}$.  \label{fig:glm} }
\end{figure}
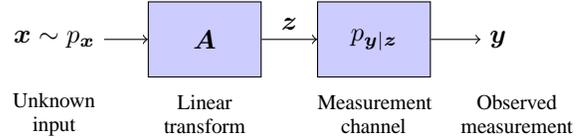

We consider the problem of estimating a random vector $\vec{x}\in\Real^N$ from observations $\vec{y}\in\Real^M$ generated as shown in \figref{glm}, which is
known as the \emph{generalized linear model} (GLM) \cite{McCullagh:Book:89}.
Under this model, $\vec{x}$ has a prior density $\pxv$ and $\vec{y}$ obeys a likelihood function of the form $p(\vec{y}|\vec{x})=\pyzv(\vec{y}|\vec{Ax})$, where $\vec{A}\in\Real^{M\times N}$ is a known linear transform and $\vec{z}\defn\vec{Ax}$ are hidden transform outputs.
The conditional density $\pyzv$ can be interpreted as a probabilistic measurement channel that accepts a vector $\vec{z}$ and outputs a random vector $\vec{y}$.
Although we have assumed real-valued quantities for the sake of simplicity, it is straightforward to generalize the methods in this paper to complex-valued quantities.

\subsection{The Generalized Linear Model} \label{sec:glm}

The GLM has many applications in statistics, computer science, and engineering.
For example, 
in \emph{statistical regression} \cite{Gelman:Book:06}, $\vec{A}$ and $\vec{y}$ contain experimental features and outcomes, respectively, and $\vec{x}$ are coefficients that best predict $\vec{y}$ from $\vec{A}$.
The relationship between $\vec{y}$ and the optimal scores $\vec{z}=\vec{Ax}$
is then characterized by $\pyzv$.
In \emph{imaging}-related inverse problems \cite{Ribes:SPM:08}, $\vec{x}$ is an image to recover, $\vec{A}$ is often Fourier-based, and $\pyzv$ models the sensor(s).
In \emph{communications} problems \cite{Hlawatsch:Book:11}, $\vec{x}$ may be a vector of discrete symbols to recover, in which case $\vec{A}$ is a function of the modulation/demodulation scheme and the propagation physics.
Or, $\vec{x}$ may contain propagation-channel parameters to recover, in which case $\vec{A}$ is a function of the modulation/demodulation scheme and the pilot symbols.
In both cases, $\pyzv$ models receiver hardware and interference.

Below we give some examples of the measurement channels $\pyzv$ that are encountered in these applications.
\begin{itemize}
\item \emph{Robust regression} \cite{Huber:Book:09} 
treats $\vec{y}=\vec{z}+\vec{w}$, and so
$\pyzv(\vec{y}|\vec{z})=\pwv(\vec{y}-\vec{z})$, where $\pwv$ is the density of $\vec{w}$.
The ``standard linear model'' treats $\vec{w}$ as additive white Gaussian noise (AWGN) but is not robust to outliers.
Robust methods use i.i.d.\ heavy-tailed models for $\vec{w}$.
\item
\emph{Binary linear classification} \cite{Bishop:Book:07} 
can be modeled using $y_m=\sgn(z_m+w_m)$, where $\sgn(v)=1$ for $v\geq 0$ and $\sgn(v)=-1$ for $v<0$, and $w_m$ are i.i.d.\ errors.
Gaussian $w_m$ yields the ``probit'' model and logistic $w_m$ yields the ``logistic'' model.
\item
\emph{Quantized compressive sensing} \cite{Kamilov:TSP:12}  
models $y_m=Q(z_m+w_m)$ with i.i.d.\ noise $w_m$. 
Here, $Q(\cdot)$ is a scalar quantizer.
\item
\emph{Phase retrieval} \cite{Millane:SPIE:06} 
uses $y_m=|z_m+w_m|$ with $z_m,w_m\in\Complex$.
When $w_m$ is i.i.d.\ circular Gaussian, $\pyzv(\vec{y}|\vec{z})=\prod_{m=1}^M\pyz(y_m|z_m)$ with Rician $\pyz(\cdot|z)$ \cite{Schniter:TSP:15}.
\item 
\emph{Photon-limited imaging} \cite{Willett:OE:11} models the number of photons collected by the sensor, $y_m$, using a Poisson distribution with rate parameter $z_m$.  
Similar models are used when inferring parameters from \emph{neural spike trains} \cite{Fletcher:NIPS:11}.
\end{itemize}

\subsection{Inference under the Generalized Linear Model} \label{sec:glm_inf}

Our goal is to estimate the random vector $\vec{x}\in\Real^N$ from the observed measurements $\vec{y}\in\Real^M$.
From the Bayesian viewpoint, there are two major options:
\emph{maximum a posteriori (MAP) estimation}
or \emph{approximate marginal inference}.
The MAP estimate is the posterior maximizer, i.e.,
\begin{align}
\hvec{x}\map
&= \arg\max_{\vec{x}} p(\vec{x}|\vec{y}) 
\stackrel{(a)}{=} \arg\max_{\vec{x}} \left\{ \ln p(\vec{y}|\vec{x}) + \ln \pxv(\vec{x}) \right\} \nonumber\\
&= \arg\max_{\vec{x}} \left\{ \ln \pyzv(\vec{y}|\vec{Ax}) + \ln \pxv(\vec{x}) \right\}
\label{eq:map} ,
\end{align}
where (a) is due to the monotonicity of the logarithm and Bayes rule, and \eqref{map} is due to the GLM.
From \eqref{map}, we see that MAP estimation is equivalent to solving an optimization problem of the form
``$\arg\min_{\vec{x}}\left\{l(\vec{x})+r(\vec{x})\right\}$,''
with loss function $l(\vec{x})\defn -\ln \pyz(\vec{y}|\vec{Ax})$
and regularizer $r(\vec{x})\defn -\ln \pxv(\vec{x})$.
Such problems are tractable when the loss and regularization are both convex.
For example, with the AWGN channel $p(\vec{y}|\vec{z})=\mc{N}(\vec{y};\vec{z},\vec{I}/\gamma_w)$ and i.i.d.\ Laplacian prior $p(x_n)=0.5\lambda\exp(-\lambda |x_n|)$, MAP estimation reduces to the LASSO \cite{Tibshirani:JRSSb:96} 
problem
``$\arg\min_{\vec{x}}\big\{\|\vec{y}-\vec{Ax}\|_2^2+\frac{\lambda}{\gamma_w} \|\vec{x}\|_1\big\}$.''

Tractable MAP optimization objectives, however, are often only surrogates for desired optimization objectives, such as minimizing the mean-squared error (MSE) on $\hvec{x}$ or the classification error rate induced by the scores $\hvec{z}=\vec{A}\hvec{x}$.
Likewise, MAP estimation returns a point estimate $\hvec{x}\map$, but reports nothing about the quality of that estimate.
Such considerations motivate a different approach, known as \emph{inference}, where the goal is to compute marginal posteriors like $p(x_n|\vec{y})$
and $p(z_m|\vec{y})$.
If $p(x_n|\vec{y})$ was known, then the minimum MSE (MMSE) estimate of $x_n$ and the MMSE itself are simply the mean and variance of $p(x_n|\vec{y})$ \cite{Poor:Book:94}.
Exact marginal inference, however, is intractable for most problems of interest. 
Thus, one must usually settle for an approximation.

One well-known approach to approximate marginal inference is through \emph{stochastic simulation} methods like MCMC \cite{Pereyra:JSTSP:16}.
But for high dimensional GLMs, such techniques can be computationally expensive and their convergence is difficult to assess. 
Another approach is \emph{variational inference} \cite{Wainwright:FTML:08}.
There, the true posterior $p(\vec{x}|\vec{y})$ is approximated by a belief $b(\vec{x})$ that is restricted to a subset of densities $\mc{Q}$ 
chosen as a compromise between fidelity and tractability.
For example, the standard ``mean field'' approach 
\cite{Parisi:Book:88} 
assumes $b(\vec{x})=\prod_{n=1}^N b_n(x_n)$ 
while the ``expectation propagation'' approach in \cite{Seeger:ECML:07} assumes $b(\vec{x})=\prod_{m=1}^M b_m(\vec{a}_m\tran\vec{x})$, where $\vec{a}_m\tran$ is the $m$th row of $\vec{A}$.
Additional constraints on the factors $b_m$ are then needed, which restricts the choice of $\pyz$ and $\px$.
Common examples include exponential-family, log-concavity, or Gaussian-scale-mixture constraints.
Furthermore, high-quality variational inference often require the inversion of an $M\!\times\! M$ or $N\!\times\! N$ matrix at each iteration, which is impractical for large $M,N$.

The \emph{approximate message passing} (AMP) algorithm \cite{Donoho:PNAS:09}, originally proposed for the \emph{standard linear model} (SLM)
\begin{align} 
\vec{y} 
&= \vec{Ax}+\vec{w} 
\text{~~with~~} 
\vec{w}\sim\mc{N}(\vec{0},\vec{I}/\gamma_w) 
\label{eq:slm},
\end{align} 
was extended to the GLM in \cite{Rangan:ISIT:11}.
The resulting \emph{generalized AMP} (GAMP) algorithm is a computationally efficient approach to either MAP or marginal inference that places few restrictions on $\pxv$ and $\pyzv$.
GAMP was originally formulated assuming a separable prior and measurement channel, i.e.,
\begin{align}
\pxv(\vec{x})=\prod_{n=1}^N \px(x_n)
\text{~~and~~}
\pyzv(\vec{y}|\vec{z})=\prod_{m=1}^M \pyz(y_m|z_m)
\label{eq:separable} ,
\end{align}
but extensions to non-identical factors and non-separable $\pxv$ and $\pyzv$ have been proposed (e.g., \cite{Schniter:CISS:10,Rangan:ISIT:12,Borgerding:ICASSP:15,Metzler:TIT:16}).
Most significantly, when $\vec{A}$ is large and i.i.d.\ zero-mean sub-Gaussian and the separability condition \eqref{separable} holds, (G)AMP is rigorously characterized by a scalar state evolution whose fixed points, when unique, are Bayes-optimal \cite{Rangan:ISIT:11,Javanmard:II:13}. 
However, (G)AMP can badly misbehave for other $\vec{A}$.
For example, small mean perturbations and/or coefficient correlations in $\vec{A}$ can cause (G)AMP to diverge \cite{Vila:ICASSP:15}.
Although damping \cite{Rangan:ISIT:14,Vila:ICASSP:15} and sequential-updating \cite{Manoel:ICML:15} strategies have been proposed to robustify (G)AMP, they are limited in their effect.

In this paper, we propose a new methodology for both MAP estimation and approximate inference under the GLM.
Our method leverages the \emph{vector AMP} (VAMP) \cite{Rangan:VAMP} framework.

\section{VAMP for the Standard Linear Model} \label{sec:vamp_slm}

We first review
the VAMP algorithm, which extends SLM-based AMP from i.i.d.\ sub-Gaussian $\vec{A}$ to ``right-rotationally invariant'' (RRI) $\vec{A}$.
RRI random matrices are described by an SVD $\vec{A}=\vec{USV}\tran$ with $\vec{V}$ uniformly distributed over the group of orthogonal matrices, allowing arbitrary deterministic $\vec{U}$ and $\vec{S}$. 
It was shown in \cite{Rangan:VAMP} that, with large RRI $\vec{A}$, VAMP can be rigorously characterized by a scalar state evolution whose fixed points agree with the replica prediction of MMSE.
Numerical experiments in \cite{Rangan:VAMP} suggest that VAMP performs very close to the replica prediction even at moderate dimensions and with strongly non-zero-mean or ill-conditioned $\vec{A}$.
Such robust behavior is not observed with the S-AMP algorithm \cite{Cakmak:ITW:14}, which enjoys the same fixed points as VAMP but does not reliably converge to those fixed points.

The VAMP algorithm for the SLM \eqref{slm} is specified in \algref{vamp_slm}.
There, $\vec{g}_1(\cdot,\gamma):\Real^N\rightarrow\Real^N$ is a ``denoising'' function 
identical to that used in the (G)AMP algorithm,
and $\bkt{\vec{g}_1'(\vec{r},\gamma)}$ is its divergence at $\vec{r}$, i.e.,
\begin{align}
\bkt{\vec{g}_i'(\vec{r},\gamma)}
&= \frac{1}{N}\tr\left\{ \frac{\partial \vec{g}_i(\vec{r},\gamma)}{\partial \vec{r} } \right\}
~~\text{for}~~ i=1,2
\label{eq:div} .
\end{align}
Under a separable prior, as in \eqref{separable}, VAMP could be configured for 
approximate marginal inference by choosing $\vec{g}_1$ as
\begin{align}
[\vec{g}_1(\vec{r},\gamma)]_n
&= \int_{\Real} \!x_n\, b(x_n;r_n,\gamma) \dif x_n
\label{eq:mmse} \\
b(x_n;r_n,\gamma) 
&\propto \px(x_n)\mc{N}(x_n;r_n,1/\gamma)
\label{eq:post} ,
\end{align}
where $b(x_n;[\vec{r}_{1k}]_n,\gamma_{1k})$ is VAMP's iteration-$k$ approximation of the marginal posterior $p(x_n|\vec{y})$.
Likewise, VAMP can be configured for MAP inference by choosing $\vec{g}_1$ as 
\begin{align}
[\vec{g}_1(\vec{r},\gamma)]_n
&=\arg\max_{x_n} b(x_n;r_n,\gamma) .
\end{align}
Non-separable priors $\pxv$ are implicitly supported by \algref{vamp_slm}, although the simpler Monte-Carlo divergence approximation from \cite[Section~V.B]{Metzler:TIT:16} has also been observed to work well in VAMP \cite{Schniter:BASP:17}.
In general, $\vec{g}_1(\cdot,\gamma)$ can be interpreted as ``denoising'' the AWGN-corrupted pseudo-measurement $\vec{r}_{1k}=\vec{x}+\mc{N}(\vec{0},\vec{I}/\gamma_{1k})$ using prior knowledge of $\vec{x}$. 

\begin{algorithm}[t]
\newcommand{\Kit}{K}
\newcommand{\kp}{k\!+\!}
\caption{VAMP for the SLM}
\begin{algorithmic}[1]  \label{alg:vamp_slm}
\REQUIRE{ LMMSE estimator
          $\vec{g}_2(\vec{r}_{2k},\gamma_{2k})$ from \eqref{g2slr_svd},
          denoiser $\vec{g}_1(\cdot,\gamma_{1k})$, 
          and number of iterations $\Kit$.  }
\STATE{ Select initial $\vec{r}_{10}$ and $\gamma_{10}\geq 0$.}
\FOR{$k=0,1,\dots,\Kit$}
    \STATE{// Denoising}
    \STATE{$\hvec{x}_{1k} = \vec{g}_1(\vec{r}_{1k},\gamma_{1k})$,~~
           $\alpha_{1k} = \bkt{ \vec{g}_1'(\vec{r}_{1k},\gamma_{1k}) }$}
        \label{line:x1}
    \STATE{$\vec{r}_{2k} = (\hvec{x}_{1k} - \alpha_{1k}\vec{r}_{1k})/(1-\alpha_{1k})$}
        \label{line:r2}
    \STATE{$\gamma_{2k} = \gamma_{1k}(1-\alpha_{1k})/\alpha_{1k}$}
        \label{line:gam2}
    \STATE{// LMMSE estimation}
    \STATE{$\hvec{x}_{2k} = \vec{g}_2(\vec{r}_{2k},\gamma_{2k})$,~~
           $\alpha_{2k} = \bkt{ \vec{g}_2'(\vec{r}_{2k},\gamma_{2k}) }$}
        \label{line:x2}
    \STATE{$\vec{r}_{1,\kp1} = (\hvec{x}_{2k} - \alpha_{2k}\vec{r}_{2k})/(1-\alpha_{2k})$}
        \label{line:r1}
    \STATE{$\gamma_{1,\kp1} = \gamma_{2k}(1-\alpha_{2k})/\alpha_{2k}$}
        \label{line:gam1}
\ENDFOR
\STATE{Return $\hvec{x}_{1\Kit}$.}
\end{algorithmic}
\end{algorithm}

The function $\vec{g}_2(\vec{r}_{2k},\gamma_{2k}):\Real^N\rightarrow\Real^N$ in \lineref{x2} of \algref{vamp_slm} performs LMMSE estimation of $\vec{x}$ from the AWGN-corrupted measurements \eqref{slm} under the pseudo-prior $\vec{x}\sim\mc{N}(\vec{r}_{2k},\vec{I}/\gamma_{2k})$, i.e.,
\begin{eqnarray}
&&\vec{g}_2(\vec{r}_{2k},\gamma_{2k})
:= \big( \gamma_w \vec{A}\tran\vec{A} + \gamma_{2k}\vec{I}\big)^{-1}
        \big( \gamma_w\vec{A}\tran\vec{y} + \gamma_{2k}\vec{r}_{2k} \big) ~\quad 
\label{eq:g2slr} \\
&&\bkt{ \vec{g}_2'(\vec{r}_{2k},\gamma_{2k}) }
= \gamma_{2k}N^{-1}\tr\big[\big(\gamma_w\vec{A}\tran\vec{A}+\gamma_{2k}\vec{I}\big)^{-1}\big] 
\label{eq:g2slrB} 
\end{eqnarray}
The per-iteration matrix inverse in \eqref{g2slr}-\eqref{g2slrB} can be avoided by precomputing the SVD $\vec{A}=\vec{USV}\tran$, after which 
\begin{align}
\vec{g}_2(\vec{r}_{2k},\gamma_{2k})
&= \vec{VD}_k \big( \tvec{y} + \gamma_{2k}\vec{V}\tran\vec{r}_{2k} \big)
\label{eq:g2slr_svd} \\
\bkt{ \vec{g}_2'(\vec{r}_{2k},\gamma_{2k}) }
&= \frac{1}{N}\sum_{n=1}^N \frac{\gamma_{2k}}{\gamma_w s_n^2 + \gamma_{2k}} ,
\end{align}
where 
$\tvec{y}=\gamma_w\vec{S}\tran\vec{U}\tran\vec{y}$
and
$\vec{D}_k$ is the $N\times N$ diagonal matrix with $[\vec{D}_k]_{nn}=(\gamma_w s_n^2+\gamma_{2k})^{-1}$.
Since $\tvec{y}$ can be precomputed, the complexity of VAMP is dominated by two matrix-vector multiplies per iteration, just like AMP.

\section{VAMP for the Generalized Linear Model} \label{sec:vamp_glm}

\algref{vamp_slm} applies VAMP to the SLM.
We now show how a small modification allows its application to the GLM.
Our approach exploits the equivalence relationship
\begin{equation}
\vec{z}=\vec{Ax}
~~\Leftrightarrow~~
\vec{0}=\mat{\vec{A} & -\vec{I}} \mat{\vec{x}\\\vec{z}} 
~\Leftrightarrow~~
\ovec{y} = \ovec{A}\ovec{x}+\ovec{w}
\label{eq:trick} ,
\end{equation}
where
$\ovec{y}\defn\vec{0}$,
$\ovec{A}\defn\mat{\vec{A}&-\vec{I}}$,
$\ovec{x}\defn\smallmat{\vec{x}\\\vec{z}}$, and
$\ovec{w}\sim\mc{N}(\vec{0},\vec{I}/\gamma_e)$ as $\gamma_e\rightarrow \infty$.
Comparing \eqref{trick} to \eqref{slm}, we see that our GLM can be expressed as an SLM where $\ovec{x}$ has two sub-vectors, the first in $\Real^N$ and the second in $\Real^M$.
Because these two sub-vectors can behave very differently, 
we propose a modified VAMP that separately tracks the precision of each.
The result, shown in \algref{vamp_glm}, can be interpreted as an instance of the more general ``GEC'' algorithm from \cite{Fletcher:ISIT:16} with a particular diagonalization operator.

In the sequel, we will use 
$\hvec{x}_{ik}\in\Real^N$ and $\hvec{z}_{ik}\in\Real^M$ to denote the two sub-vectors of the output of $\vec{g}_i$ at iteration $k$ (for $i=1,2$), 
and we will use 
$\vec{r}_{ik}\in\Real^N$ and $\vec{p}_{ik}\in\Real^M$ to denote the two sub-vectors of the input to $\vec{g}_i$.
As in SLM-based VAMP, 
we will use the pseudo-measurement model 
$\vec{r}_{1k}=\vec{x}+\mc{N}(\vec{0},\vec{I}/\gamma_{1k})$ when denoising $\vec{x}$
and the pseudo-prior
$\vec{x}\sim\mc{N}(\vec{r}_{2k},\vec{I}/\gamma_{2k})$ for LMMSE estimation of $\vec{x}$.
Likewise, we will use pseudo-measurements 
$\vec{p}_{1k}=\vec{z}+\mc{N}(\vec{0},\vec{I}/\tau_{1k})$ when denoising $\vec{z}$
and the pseudo-prior
$\vec{z}\sim\mc{N}(\vec{p}_{2k},\vec{I}/\tau_{2k})$ for LMMSE estimation of $\vec{z}$.
A rigorous justification of these models is postponed for future work.

The independence between the random variables $\vec{x}$ and the random variables $\vec{y}$ conditioned on $\vec{z}$ implies that the function $\vec{g}_1$ decouples across the two sub-vectors.
That is, we can write 
$\hvec{x}_{1k}=\vec{g}_{x1}(\vec{r}_{1k},\gamma_{1k})$ and
$\hvec{z}_{1k}=\vec{g}_{z1}(\vec{p}_{1k},\tau_{1k})$
for denoisers $\vec{g}_{x1}(\cdot,\gamma_{1k}):\Real^N\rightarrow\Real^N$ 
and $\vec{g}_{z1}(\cdot,\tau_{1k}):\Real^M\rightarrow\Real^M$.
The construction of $\vec{g}_{x1}$ remains the same as described in \secref{vamp_slm}, and the construction of $\vec{g}_{z2}$ is similar but with $\pyzv(\vec{y}|\cdot)$ replacing $\pxv(\cdot)$.
Lines~\ref{line:r2g}-\ref{line:gam2g} and \ref{line:p2g}-\ref{line:tau2g} of \algref{vamp_glm} follow directly from lines~\ref{line:r2}-\ref{line:gam2} of \algref{vamp_slm}.

\begin{algorithm}[t]
\newcommand{\Kit}{K}
\newcommand{\kp}{k\!+\!}
\caption{VAMP for the GLM}
\begin{algorithmic}[1]  \label{alg:vamp_glm}
\REQUIRE{ LMMSE estimators $\vec{g}_{x2}$ and $\vec{g}_{z2}$ 
          from \eqref{g2glm_svdA} or \eqref{g2glm_svdB},
          denoisers $\vec{g}_{x1}$ and $\vec{g}_{z1}$,
          and number of iterations $\Kit$. }
\STATE{ Select initial $\vec{r}_{10},\vec{p}_{10},\gamma_{10}>0,\tau_{10}>0$.}
\FOR{$k=0,1,\dots,\Kit$}
    \STATE{// Denoising $\vec{x}$ }
    \STATE{$\hvec{x}_{1k} = \vec{g}_{x1}(\vec{r}_{1k},\gamma_{1k})$,~~
           $\alpha_{1k} = \bkt{ \vec{g}_{x1}'(\vec{r}_{1k},\gamma_{1k}) }$}
        \label{line:x1g}
    \STATE{$\vec{r}_{2k} = (\hvec{x}_{1k} - \alpha_{1k}\vec{r}_{1k})/(1-\alpha_{1k})$}
        \label{line:r2g}
    \STATE{$\gamma_{2k} = \gamma_{1k}(1-\alpha_{1k})/\alpha_{1k}$}
        \label{line:gam2g}
    \STATE{// Denoising $\vec{z}$}
    \STATE{$\hvec{z}_{1k} = \vec{g}_{z1}(\vec{p}_{1k},\tau_{1k})$,~~
           $\beta_{1k} = \bkt{ \vec{g}_{z1}'(\vec{p}_{1k},\tau_{1k}) }$}
        \label{line:z1g}
    \STATE{$\vec{p}_{2k} = (\hvec{z}_{1k} - \beta_{1k}\vec{p}_{1k})/(1-\beta_{1k})$}
        \label{line:p2g}
    \STATE{$\tau_{2k} = \tau_{1k}(1-\beta_{1k})/\beta_{1k}$}
        \label{line:tau2g}
    \STATE{// LMMSE estimation of $\vec{x}$}
    \STATE{$\hvec{x}_{2k} = \vec{g}_{x2}(\vec{r}_{2k},\vec{p}_{2k},\gamma_{2k},\tau_{2k})$,~~
           $\alpha_{2k} = \bkt{\vec{g}_{x2}'(\dots)}$}
        \label{line:x2g}
    \STATE{$\vec{r}_{1,\kp1} = (\hvec{x}_{2k} - \alpha_{2k}\vec{r}_{2k})/(1-\alpha_{2k})$}
        \label{line:r1g}
    \STATE{$\gamma_{1,\kp1} = \gamma_{2k}(1-\alpha_{2k})/\alpha_{2k}$}
        \label{line:gam1g}
    \STATE{// LMMSE estimation of $\vec{z}$}
    \STATE{$\hvec{z}_{2k} = \vec{g}_{z2}(\vec{r}_{2k},\vec{p}_{2k},\gamma_{2k},\tau_{2k})$,~~
           $\beta_{2k} = \bkt{\vec{g}_{z2}'(\dots)}$}
        \label{line:z2g}
    \STATE{$\vec{p}_{1,\kp1} = (\hvec{z}_{2k} - \beta_{2k}\vec{p}_{2k})/(1-\beta_{2k})$}
        \label{line:p1g}
    \STATE{$\tau_{1,\kp1} = \tau_{2k}(1-\beta_{2k})/\beta_{2k}$}
        \label{line:tau1g}
\ENDFOR
\STATE{Return $\hvec{x}_{1\Kit}$.}
\end{algorithmic}
\end{algorithm}

Lines~\ref{line:x2g}-\ref{line:tau1g} of \algref{vamp_glm} implement LMMSE estimation of $\ovec{x}=\smallmat{\vec{x}\\\vec{z}}$ under the SLM in \eqref{trick} and the pseudo-prior
\begin{equation}
\ovec{x} =
\mat{\vec{x}\\\vec{z}}
\sim\mc{N}\left(\mat{\vec{r}_{2k}\\\vec{p}_{2k}},\mat{\vec{I}/\gamma_{2k}\\&\vec{I}/\tau_{2k}}\right) .
\end{equation}
Because the likelihood and prior are both Gaussian, the LMMSE estimate is equivalent to the MAP estimate 
\begin{align}
\lefteqn{
\arg\max_{\ovec{x}} p(\ovec{x}|\ovec{y})
=\arg\min_{\ovec{x}} \left\{ -\ln p(\ovec{y}|\ovec{x}) -\ln p(\ovec{x}) \right\} }
\label{eq:g2glmA}\\
&=\arg\min_{\vec{x},\vec{z}} \gamma_e\|\vec{Ax}-\vec{z}\|_2^2
        + \gamma_{2k} \|\vec{r}_{2k}-\vec{x}\|_2^2 + \tau_{2k} \|\vec{p}_{2k}-\vec{z}\|_2^2 
\nonumber .
\end{align}
Zeroing the gradients w.r.t.\ $\vec{x}$ and $\vec{z}$, taking $\gamma_e\rightarrow \infty$, and substituting the SVD $\vec{A}=\vec{USV}\tran$ into the result, we get 
\begin{eqnarray}
\vec{g}_{x2}(\vec{r}_{2k},\vec{p}_{2k},\gamma_{2k},\tau_{2k})
&=& \vec{V}\vec{D}_k \big( \tau_{2k}\vec{S}\tran\vec{U}\tran\vec{p}_{2k} + \gamma_{2k}\vec{V}\tran\vec{r}_{2k} \big)  
\nonumber \\
\vec{g}_{z2}(\vec{r}_{2k},\vec{p}_{2k},\gamma_{2k},\tau_{2k})
&=& \vec{A}\vec{g}_{x2}(\vec{r}_{2k},\vec{p}_{2k},\gamma_{2k},\tau_{2k}) 
\label{eq:g2glm_svdA} ,
\end{eqnarray}
where $\vec{D}_k$ is an $N\times N$ diagonal matrix such that 
$[\vec{D}_k]_{nn} \defn (\tau_{2k}s_n^2+\gamma_{2k})^{-1}$.
An alternative expression for $\vec{g}_{x2}$ is
\begin{eqnarray}
\lefteqn{ \vec{g}_{x2}(\vec{r}_{2k},\vec{p}_{2k},\gamma_{2k},\tau_{2k}) }\nonumber\\
&=& \vec{r}_{2k} + \vec{VS}\tran\Big( \frac{\gamma_{2k}}{\tau_{2k}}\vec{I}+\vec{SS}\tran\Big)^{-1}\big(\vec{U}\tran\vec{p}_{2k}-\vec{SV}\tran\vec{r}_{2k}\big)
. \qquad \label{eq:g2glm_svdB} 
\end{eqnarray}
Both \eqref{g2glm_svdA} and \eqref{g2glm_svdB} are derived in the Appendix.

Recalling the definition of the divergence in \eqref{div}, we see that 
$\alpha_{2k}$ from \lineref{x2g} of \algref{vamp_glm} 
equals $N^{-1}$ times the trace of the Jacobian $\partial \vec{g}_{x2}/\partial \vec{r}_{2k}=\gamma_{2k}\vec{VD}_k\vec{V}\tran$, and so \eqref{g2glm_svdB} gives
\begin{eqnarray}
\alpha_{2k} 
= \bkt{ \vec{g}_{x2}'(\vec{r}_{2k},\vec{p}_{2k},\gamma_{2k},\tau_{2k}) }
&=& \frac{1}{N}\sum_{n=1}^N \frac{\gamma_{2k}}{\tau_{2k} s_n^2 + \gamma_{2k}} .
\qquad \label{eq:a2k}
\end{eqnarray}
Similarly, $\beta_{2k}$ from \lineref{z2g} of \algref{vamp_glm}
is $M^{-1}$ times the trace of the Jacobian $\partial \vec{g}_{z2}/\partial \vec{p}_{2k}=\tau_{2k}\vec{SD}_k\vec{S}\tran$, and so 
\begin{align}
\beta_{2k}
&= \bkt{ \vec{g}_{z2}'(\vec{r}_{2k},\vec{p}_{2k},\gamma_{2k},\tau_{2k}) } \\
&= \frac{1}{M}\sum_{n=1}^N \frac{\tau_{2k} s_n^2}{\tau_{2k} s_n^2 + \gamma_{2k}}
= \frac{M}{N}(1-\alpha_{2k}) 
\label{eq:b2k} .
\end{align}
The above explains lines~\ref{line:x2g}~and~\ref{line:z2g} of \algref{vamp_glm}.
Lines~\ref{line:r1g}-\ref{line:gam1g} and \ref{line:p1g}-\ref{line:tau1g} of \algref{vamp_glm} follow directly from 
lines~\ref{line:r1}-\ref{line:gam1} of \algref{vamp_slm}.

%

\section{Numerical Experiments}

We now show the results of a numerical experiment on \emph{one-bit compressed sensing}, where the goal was to recover the sparse signal $\vec{x}\in\Real^N$ from measurements 
\begin{align}
y_m
&= \sgn\big([\vec{Ax} + \vec{w}]_m\big)
~~\text{for}~~m=1,\dots,M
\label{eq:y} .
\end{align}
For our experiment, 
we drew $\vec{w}\sim\mc{N}(\vec{0},\vec{I}/\gamma_w)$ 
and we constructed $\vec{x}$ with $16$ non-zero coefficients 
whose amplitudes were drawn i.i.d.\ $\mc{N}(0,1)$ and 
whose indices were drawn independently and uniformly at random.
Also, we used $N=512$ and $M=2048$,
and we adjusted $\gamma_w$ to achieve a signal-to-noise ratio $\E\{\|\vec{Ax}\|^2\}/\E\{\|\vec{w}\|^2\}=40$~dB.
 
Following \cite{Vila:ICASSP:15}, we constructed $\vec{A} \in \Real^{M \times N}$ from the singular value decomposition (SVD) $\vec{A}=\vec{U}\vec{S}\vec{V}\tran$, where orthogonal matrices $\vec{U}$ and $\vec{V}$ were drawn uniformly with respect to the Haar measure.
That is, $\vec{A}$ was rotationally invariant.
The singular values $s_n$ were a geometric series, i.e., $s_n/s_{n-1}=\rho~\forall n>1$, with $\rho$ and $s_1$ chosen to achieve a desired condition number $\kappa(\vec{A})\defn s_1/s_{\min(M,N)}$ with $\|\vec{A}\|_F^2=N$.
It was shown in \cite{Rangan:ISIT:14,Vila:ICASSP:15} that standard AMP
(and even damped AMP) diverges when the matrix $\vec{A}$ has a sufficiently high condition number.
Thus, this matrix-generation model provides an effective test for the stability of AMP methods.
Recovery performance was assessed using ``debiased'' normalized mean-squared error (dNMSE), $\min_{c\in\Real}\|c \hvec{x}-\vec{x}\|^2/\|\vec{x}\|^2$. 
The debiasing was used because the measurement channel discards amplitude information.

\Figref{probit_vs_cond} plots the average dNMSE achieved by VAMP and by the adaptively damped (AD) GAMP algorithm from \cite{Vila:ICASSP:15} versus condition number $\kappa(\vec{A})$.
The dNMSE was evaluated for $\kappa(\vec{A})$ ranging from $1$ (i.e., row-orthogonal) to $10^6$ (i.e., highly ill-conditioned $\vec{A}$), and averaged
over $500$ independent draws of $\vec{A}$, $\vec{x}$, and $\vec{w}$.
For this experiment, VAMP perfectly knew the prior $\pxv$ and measurement-channel $\pyzv$ (although if not the technique in \cite{Fletcher:EMVAMP} could be used for automatic tuning) 
and it was initialized using $\vec{r}_{10}=\vec{0}$, $\vec{p}_{10}=\vec{0}$, $\gamma_{10}=10^{-8}$, and $\tau_{10}=10^{-8}$.
The figure shows that AD-GAMP accurately recovered $\vec{x}$ 
for $\kappa(\vec{A})<10^3$ but failed at higher condition numbers.
By contrast, VAMP accurately recovered $\vec{x}$ over the full tested range of $\kappa(\vec{A})$.

\Figref{probit_vs_iter} plots the average dNMSE versus iteration for condition numbers $\kappa(\vec{A})\in\{1,316,10^6\}$.
The figures show that, for the range of $\kappa(\vec{A})$ where AD-GAMP accurately recovers $\vec{x}$, VAMP converges faster: in about $10$ iterations compared to $30$-$40$ for AD-GAMP.
Meanwhile, at the extreme case of $\kappa(\vec{A})=10^6$, VAMP converges in less than $20$ iterations. 
Thus, these experiments suggest that the convergence speed of VAMP is relatively insensitive to the condition number of large, rotationally invariant $\vec{A}$.

\putFrag{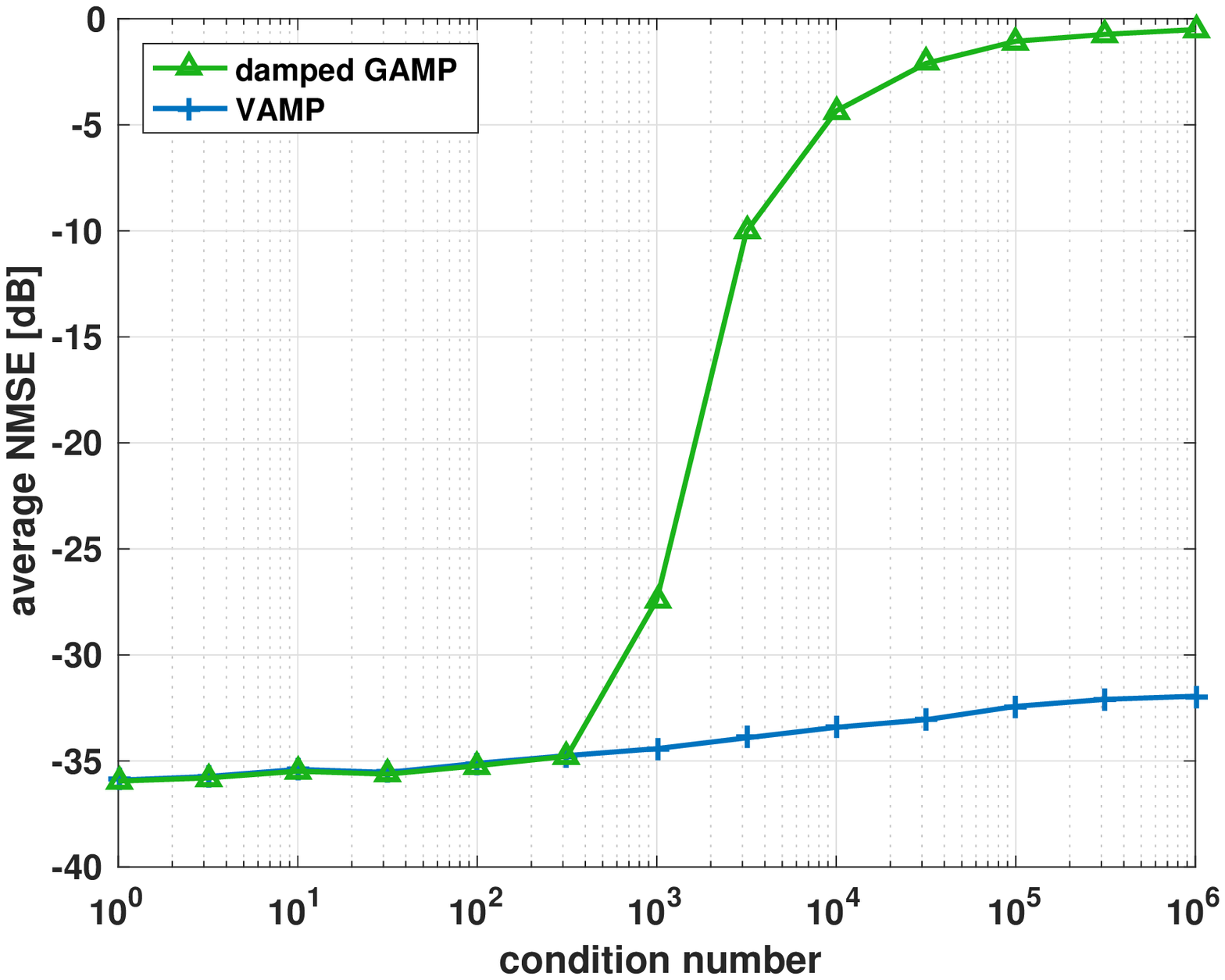}
        {Debiased NMSE versus condition number $\kappa(\vec{A})$ at the final algorithm iteration, averaged over $500$ realizations.}
        {\figsize}
        {\newcommand{\sz}{0.8}
         \newcommand{\szz}{0.5}
         \psfrag{condition number}[t][t][\sz]{\sf condition number} 
         \psfrag{average NMSE [dB]}[b][b][\sz]{\sf average dNMSE [dB]}
         \psfrag{damped GAMP}[Bl][Bl][\szz]{\sf AD-GAMP} 
         \psfrag{VAMP}[Bl][Bl][\szz]{\sf VAMP} 
         \psfrag{Probit, SNR=40dB, N=512, M=2048, K=16, U=Haar, V=Haar, mean of 500}{}}

\putFrag{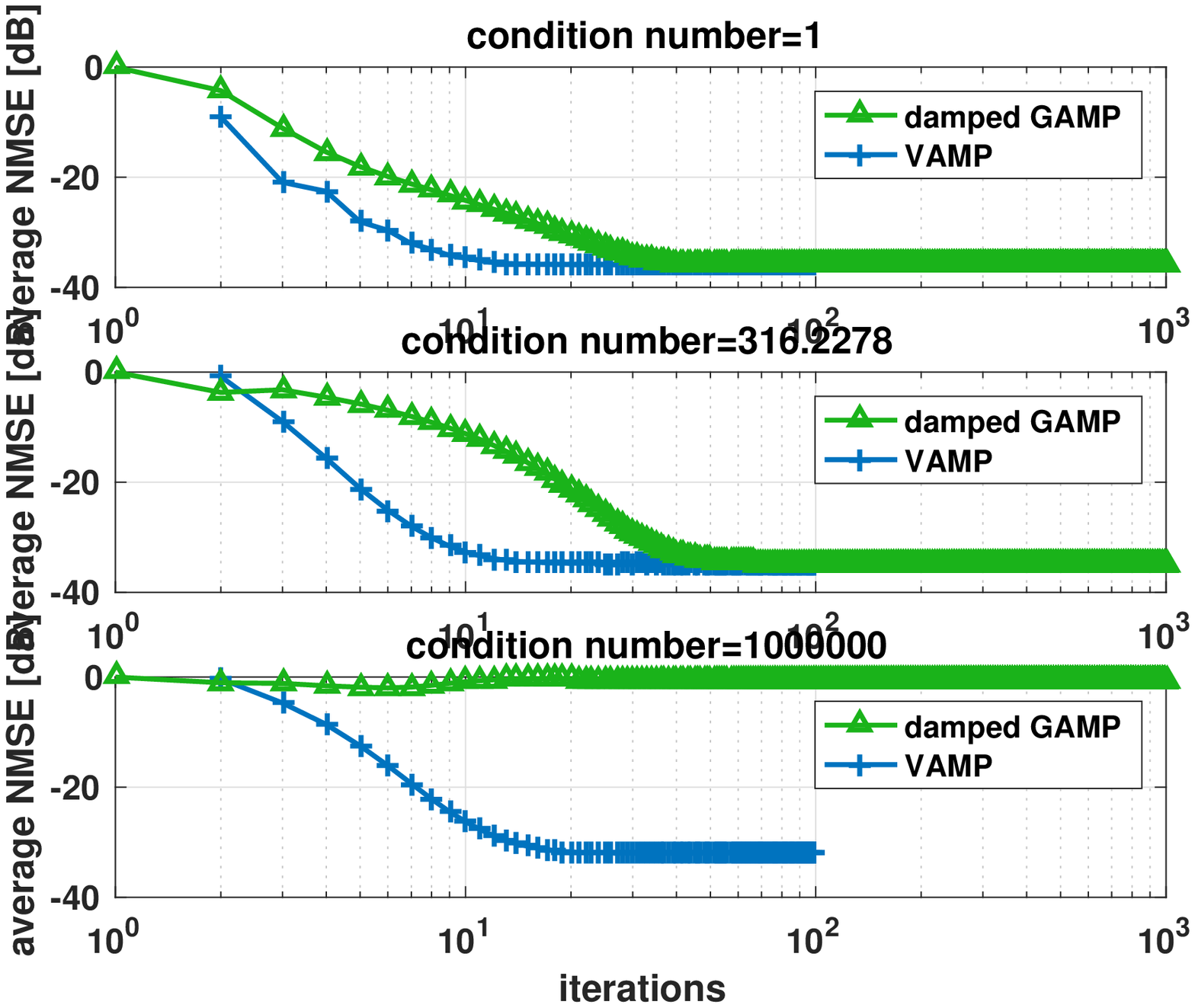}
        {Debiased NMSE versus iteration $k$ at several condition numbers $\kappa(\vec{A})=1$ in (a), $\kappa(\vec{A})=316.23$ in (b), and $\kappa(\vec{A})=10^6$ in (c), averaged over $500$ realizations.}
        {\figsize}
        {\newcommand{\sz}{0.8}
         \newcommand{\szz}{0.5}
         \psfrag{iterations}[t][t][\sz]{\sf iterations} 
         \psfrag{average NMSE [dB]}[b][b][\sz]{\sf dNMSE [dB]}
         \psfrag{damped GAMP}[Bl][Bl][\szz]{\sf AD-GAMP} 
         \psfrag{VAMP}[Bl][Bl][\szz]{\sf VAMP} 
         \psfrag{condition number=1}[B][b][\sz]{\sf $\kappa(\vec{A})=1$} 
         \psfrag{condition number=316.2278}[B][b][\sz]{\sf $\kappa(\vec{A})=316$} 
         \psfrag{condition number=1000000}[B][b][\sz]{\sf $\kappa(\vec{A})=10^6$} 
         \psfrag{}{}}

\appendix

To derive \eqref{g2glm_svdA}-\eqref{g2glm_svdB}, we
zero the gradient of the cost in \eqref{g2glmA} w.r.t.\ $\vec{x}$ and $\vec{z}$ at $\hvec{x}_{2k}$ and $\hvec{z}_{2k}$, yielding the equations
\begin{align}
\vec{0} 
&= \gamma_e \vec{A}\tran(\vec{A}\hvec{x}_{2k}-\hvec{z}_{2k}) + \gamma_{2k}(\hvec{x}_{2k}-\vec{r}_{2k}) \\
\vec{0} 
&= \gamma_e (\hvec{z}_{2k}-\vec{A}\hvec{x}_{2k}) + \tau_{2k}(\hvec{z}_{2k}-\vec{p}_{2k}) ,
\end{align}
which can be rewritten as
\begin{align}
\mat{\gamma_{2k}\vec{r}_{2k}\\\tau_{2k}\vec{p}_{2k}}
&= \mat{\gamma_e\vec{A}\tran\vec{A}+\gamma_{2k}\vec{I} 
        & -\gamma_e \vec{A}\tran\\
        -\gamma_e \vec{A} 
        & (\tau_{2k}+\gamma_e)\vec{I} }
   \mat{\hvec{x}_{2k}\\\hvec{z}_{2k}} 
\label{eq:g2glmB} .
\end{align}
Inverting the block matrix in \eqref{g2glmB} via the Schur complement
$\vec{Q}\defn \gamma_e\vec{A}\tran\vec{A}+\gamma_{2k}\vec{I}-\frac{\gamma_e^2}{\tau_{2k}+\gamma_e}\vec{A}\tran\vec{A}
= \frac{\gamma_e\tau_{2k}}{\tau_{2k}+\gamma_e}\vec{A}\tran\vec{A}+\gamma_{2k}\vec{I}$
gives (after temporarily suppressing the ``$k$'' index)
\begin{equation*}
\mat{\hvec{x}_{2}\\\hvec{z}_{2}} 
= \mat{\vec{Q}^{-1} 
        & \frac{\gamma_e}{\tau_{2}+\gamma_e} \vec{Q}^{-1}\vec{A}\tran \\
        \frac{\gamma_e}{\tau_{2}+\gamma_e} \vec{A}\vec{Q}^{-1}
        & \frac{1}{\tau_{2}+\gamma_e}(\vec{I}+
        \frac{\gamma_e^2}{\tau_{2}+\gamma_e} \vec{A}\vec{Q}^{-1}\vec{A}\tran) 
        }
\mat{\gamma_{2}\vec{r}_{2}\\\tau_{2}\vec{p}_{2}} .
\end{equation*}
Taking $\gamma_e\rightarrow\infty$ then gives
$\vec{Q}=\tau_{2k}\vec{A}\tran\vec{A}+\gamma_{2k}\vec{I}$ and
\begin{align}
\lefteqn{
\mat{\hvec{x}_{2k}\\\hvec{z}_{2k}} 
= \mat{\vec{Q}^{-1} 
        & \vec{Q}^{-1}\vec{A}\tran \\
        \vec{A}\vec{Q}^{-1}
        & \vec{A}\vec{Q}^{-1}\vec{A}\tran }
\mat{\gamma_{2k}\vec{r}_{2k}\\\tau_{2k}\vec{p}_{2k}} }\\
&= \mat{\vec{I}\\\vec{A}}
   \big(\tau_{2k}\vec{A}\tran\vec{A}+\gamma_{2k}\vec{I}\big)^{-1}
   \big(\gamma_{2k}\vec{r}_{2k}+\tau_{2k}\vec{A}\tran\vec{p}_{2k}\big) 
\label{eq:g2glmC} .
\end{align}
Plugging the SVD $\vec{A}=\vec{USV}\tran$ into \eqref{g2glmC} yields
\eqref{g2glm_svdA}.
An alternative expression results from the matrix inversion lemma: 
\begin{eqnarray}
\hvec{x}_{2k}
= \vec{r}_{2k} + \vec{A}\tran\Big( \frac{\gamma_{2k}}{\tau_{2k}}\vec{I}+\vec{AA}\tran\Big)^{-1}(\vec{p}_{2k}-\vec{Ar}_{2k})
\label{eq:g2glmD} ,
\end{eqnarray}
and plugging the SVD $\vec{A}=\vec{USV}\tran$ into \eqref{g2glmD} yields \eqref{g2glm_svdB}.

\bibliographystyle{ieeetr}
\bibliography{macros_abbrev,stc,books,misc,sparse,machine,phase}

\end{document}